\documentclass[twocolumn]{aastex62}

\usepackage[utf8]{inputenc}
\usepackage[draft]{fixme}
\usepackage{graphicx,xspace}
\usepackage{hyperref}
\usepackage{natbib  }
\usepackage{breakurl}
\usepackage{soul}
\usepackage{amsthm}
\usepackage{amsmath}

\begin{document}
	\title{Probing the IGMF with the next generation Cherenkov telescopes}

\correspondingauthor{M. Fernández Alonso}
\email{mateofa@iafe.uba.ar}	
	
	\author{M. Fernández Alonso}
	\author{A. D. Supanitsky}
	\author{A. C. Rovero}

\affil{Instituto de Astronom\'{\i}a y  F\'{\i}sica del Espacio (IAFE, CONICET$-$UBA), CC 67, Suc. 28, 1428 Buenos Aires, Argentina}

	\begin{abstract}

Intergalactic space is believed to contain non-zero magnetic fields (the Intergalactic Magnetic Field: IGMF) which at scales of Mpc would have intensities below $10^{-9}$ G. Very high energy (VHE $>$100 GeV) gamma rays coming from blazars can produce e$^+$e$^-$ pairs when interacting with the Extragalactic Background Light (EBL) and the Cosmic Microwave Background, generating an electromagnetic cascade of Mpc scale. The IGMF may produce a detectable broadening of the emission beam that could lead to important constrains both on the IGMF intensity and its coherence length. Using the Monte Carlo-based Elmag code, we simulate the electromagnetic cascade corresponding to two detected TeV sources: PKS 2155-304 visible from the South and H1426+428 visible from the North. Assuming an EBL model and intrinsic spectral properties of the sources we obtain the spectral and angular distribution of photons when they arrive at Earth. We include the response of the next generation Cherenkov telescopes by using simplified models for CTA (Cherenkov Telescope Array)-south and CTA-north based on a full simulation of each array performance.  Combining the instrument properties with the simulated source fluxes, we calculate the telescope point spread function for null and non-null IGMF intensities and develop a method to test the statistical feasibility of detecting IGMF imprints by comparing the resulting angular distributions. Our results show that for the analysed source PKS 2155-304 corresponding to the southern site, CTA should be able to detect IGMF with intensities stronger than 10$^{-14.5}$G within an observation time of $\sim$100 hours. 
	\end{abstract}


\section{Introduction}

Intergalactic space is believed to contain non-zero magnetic fields (the Intergalactic Magnetic Field: IGMF). It has been suggested that it could be originated in the early Universe during the electroweak or QCD phase transition \citep{2013A&ARv..21...62D}. A different proposed explanation suggests this primordial magnetic field could have been originated during the early formation of large scale structures at redshifts z $\leq$ 10 \citep{2006MNRAS.370..319B}. To the date, we have no certain information about the IGMF intensity and spatial properties and there is no direct way of probing it with present techniques. However, in the last 10 years several constrains to these parameters were derived indirectly using different methods and techniques. The non-observation of Faraday rotation induced by an IGMF in quasar observations suggests that its intensity is weaker than $10^{-9}$ G for typical Mpc scale coherence lengths (\citealp{1999ApJ...514L..79B};  \citealp{2015MNRAS.452.2851P}). Gamma-ray observations from distant active galactic nuclei (AGN) allowed the estimation of lower limits and other constrains to the parameter space of the IGMF.  An exclusion region in the range (0.3-3)$\times$10$^{-15}$G was derived using HESS blazar TeV observations assuming a Mpc scale IGMF \citep{Abramow2014A&A...562A.145H}. Similarly, an exclusion region beteween 5.5$\times$10$^{-15}$G and 7.4$\times$10$^{-14}$G was calculated by VERITAS also using blazar TeV observations and Mpc scale coherence length assumptions \citep{2017ApJ...835..288A}. Fermi-LAT observations in the GeV range also allowed the exclusion of fields below $\sim$10$^{-19}$G for coherent lengths of $>$1 Mpc \citep{Finke2015ApJ...814...20F} and below 3$\times$10$^{-16}$G for coherence lengths of $\geq$10 kpc \cite{2018ApJS..237...32A}.

Another study conducted by \cite{Arlen2014ApJ...796...18A} makes a revision of other publications where they use methods to derive lower limits, and claims that a zero-IGMF hypothesis cannot be discarded with the available data. There is still a wide range of possible values for the spatial properties and intensity of the IGMF, gamma-ray interactions in the intergalactic medium could help constrain this parameter space and even detect the IGMF indirectly.

The Universe is opaque for gamma rays in the VHE ($>$100 GeV) range. Photon absorption in the intergalactic (IG)  photon backgrounds is energy dependent and starts to become substantial at TeV energies \citep{1966PhRvL..16..252G}. In particular, VHE gamma rays from jets of AGN can interact with photons in the IR-UV range present in the Extragalactic Background Light (EBL) and photons from the Cosmic Microwave Background (CMB), producing electron-positron pairs. These pairs carry most of the energy from the original photons, and can interact as well with IG photons from the backgrounds via Inverse Compton, promoting them to energies in the HE ($>$100 MeV)-VHE range, and making them capable to pair produce in the IG backgrounds again. This cascade process converts the initial VHE photons into photons of lower energies which can travel further. Moreover, depending on the intensity ($B$) of the IGMF, the bending effect on the electron-positron pair trajectories can result into different emission scenarios. For a {\it strong} IGMF intensity ($B >10^{-7}$ G) synchrotron cooling would become dominant and no secondary gamma rays would be produced \citep{1978ApJ...225..318G}, however, as mentioned above, this scenario has been ruled out for Mpc scale IGMF by the non-observation of Faraday rotation. For a {\it moderate} IGMF ($10^{-12}$ G $< B <10^{-7}$ G) the electron and positron pair trajectories are isotropized around the source eventually giving rise to an extended isotropic emission of photons, or {\it halo}, which take much longer to reach the observer than the direct photons from the source \citep{1994ApJ...423L...5A}. For a {\it weak} IGMF ($B <10^{-14}$ G) the cascade develops almost exclusively in the forward direction, although there is a broadening of the original emission beam, even for very small IGMF intensities. 

The extension of this emission depends on the IGMF intensity, its coherence length, and the source distance and should be clearly distinguished from the halo emission because in this case the broadening takes place along the jet direction, not in an isotropic way (\citealp{Ahlers2011PhRvD..84f3006A}; \citealp{Abramow2014A&A...562A.145H}). Different assumptions for the coherence length are present in the literature, ranging from 10$^{-4}$ to 10$^4$ Mpc. The general trend is that for relatively low coherent lengths $<$ 1 Mpc, weak and moderate ($<$ 10$^{-15}$ G) IGMF intensities are ruled out. This is a result of the random change in direction of the e$^{+-}$ pairs as they cross multiple coherent lengths. For relatively high coherent lengths $>$ 1 Mpc, the intensity and coherence length are practically independent and almost all IGMF intensity scenarios are allowed \citep{Finke2015ApJ...814...20F}. 
Assumptions on the Doppler factor ($\Gamma$) and the opening angle associated to the emission jet may also play an important role in dimming or enhancing the resulting secondary radiation. Although some important effects are expected in the HE part of the energy spectrum for relatively high $\Gamma$ values ($\sim$10$^4$-10$^5$), for reasonable low values of $\Gamma<$ 100, effects in the VHE part of the spectrum can be considered negligible \citep{Arlen2014ApJ...796...18A}.

Since this effect was proposed several groups have tried unsuccessfully to observe it in the TeV band using multiple methods: \cite{2001A&A...366..746A,2010A&A...524A..77A,2010tsra.confE.192F,2014arXiv1406.4764F,Abramow2014A&A...562A.145H,Caprini2015PhRvD..91l3514C}. Other authors found evidence of extended emission around extragalactic sources in the GeV range using Fermi-LAT observations (\citealp{Chen2015PhRvL.115u1103C}; \citealp{2015MNRAS.450L..44K}). This extension could be potentially caused by the IGMF.
All these studies were done using blazars, a subtype of AGN that have their jets pointing towards the Earth and are therefore extremely luminous objects in the TeV band, perfect candidates to perform this type of studies. Since IGMF presence in the intergalactic medium will presumably affect the observed spectral and angular distributions of gamma rays coming from blazars, methods usually consist in putting these distributions under a thorough analysis (\citealp{Neronov2009PhRvD..80l3012N}; \citealp{Aharonian2010PhRvD..82d3002A}). 

The intergalactic cascade process is usually understood under the assumption that inverse Compton is the primary mechanism for the energy loss of the charged particles within the cascade. In a study,  \citet{2012ApJ...752...22B} question this idea and suggest that for bright sources ($\geq$ 10$^{42}$erg s$^{-1}$), plasma instabilities could be the main mechanism for energy loss of the produced pairs. In that case, the energy of the pairs would end up heating the intergalactic medium instead of scattering CMB or EBL photons, and no cascade process is produced.

In this work we use MC based simulations of intergalactic cascades under different IGMF scenarios, and quantitatively study the effects of the magnetic field on the resulting spectral and angular distributions of the arriving photons. Motivated by the existence of a real future Cherenkov telescope system, the Cherenkov Telescope Array (CTA) \citep{CTA_Concept}, we assume a simplified model of response for CTA-south and CTA-north and develop a method for testing the feasibility of detecting an extended component within the angular distribution of photons for each IGMF scenario. The resulting method constitutes an alternative way of studying the IGMF with the next generation Cherenkov telescopes, different from previous approaches usually based on possible IGMF imprints in the spectral energy distributions of VHE sources \citep{Meyer2016ApJ...827..147M}. A preliminary discussion about the basis of this method was presented in \cite{Fernandez2015ICRC...34..791F}.

%The detection of the halo, or the magnetic broadening might lead to important constrains on both the IGMF intensity and the EBL density, quantities that currently remain uncertain and are of great importance in cosmological models.\\
%The two key quantities are the intensity $B$ and coherence length $\lambda$, since these two may give clues about the origins of the magnetic field \citep{2013A&ARv..21...62D}.

%Very high energy (VHE) gamma rays from jets of active galactic nuclei (AGN) can interact with photons in the IR-UV range present in the Extragalactic Background Light (EBL), producing electron-positron pairs. These pairs carry most of the energy from the original photons, and can interact as well with CMB photons via Inverse Compton scattering, promoting them to energies in the HE ($>$100 MeV) to VHE range and making them able to pair produce in the inter galactic backgrounds again. This cascade process converts the initial VHE photons into photons of lower energy which can travel further \citep{1994ApJ...423L...5A}. Moreover, depending on the intensity and correlation length of the IGMF, the bending effect on the electron-positron pair trajectories results in a different spatial development of the cascade. In this way, the energy distribution and angular and temporal properties of the cascade emission coming from blazars can provide observable signatures for the IGMF.

\section{Intrument Response}
\label{Inst_response}
Current imaging atmospheric Cherenkov telescopes (IACT) like HESS, MAGIC and VERITAS have not been able to positively detect IGMF effects in TeV observations yet. The IGMF imprints are in principle present in both the spectral energy and the angular distributions, but are estimated to be relatively small. Instruments capable of detecting femto-Gauss IGMF effects are expected to have a considerable improvement in sensitivity as well as angular resolution in comparison with present instruments. CTA represents the next generation of Cherenkov Telecopes and will consist of two arrays covering a wider collection area and energy range than ever before for this kind of instruments. CTA is designed to increase significantly the effective area and the angular resolution as well as to improve the sensitivity in about an order of magnitude in relation with present instruments, making it a promising instrument for studying extended emission in the near future \citep{CTA_Concept}. For this study we consider a simplified model for the CTA array performance described in \cite{2016APh....80...22A} and combine it with CTA-north and CTA-south public performance files to model each array separately \citep{CTAperf}. In particular, we use analytic descriptions for the effective area, angular resolution and point spread function (PSF) shape to estimate the possible response of the telescopes to blazar observations.
%footnote{https://portal.cta-observatory.org/CTA_Observatory/performance/SitePages/Home.aspx}

\subsection{Effective area and Sensitivity} 

The effective area of an array of Cherenkov telescopes is determined by the total geometrical area covered by the array and the energy of the gamma-ray photon generating the light pool \citep{1997APh.....6..369A}. The effective area for one possible CTA layout can be parametrized between 50 GeV and 100 TeV by the following expression \citep{2016APh....80...22A},
\begin{equation}
A_{eff}(x)=\frac{A}{1+B\exp\left( {-\frac{x}{C}}\right)}.
\end{equation}
Here $x=\log(E/\textrm{TeV})$ where $E$ is the gamma-ray energy. The corresponding parameters for the southern and northern sites are shown in Table \ref{table_effA}.
\begin{table}[!t]
\centering
\caption{Parameters of the fitted effective area for both sites.}

\begin{tabular}{cccc}
\hline\hline\noalign{\smallskip}
\!\! Site & \!\!\!\! A [m$^2$] & \!\!\!\! B  & \!\!\!\! C \\

\hline\noalign{\smallskip}
\!\! South  &  $4.36\times10^{6}$ & 6.05 & 0.399 \\
\!\! North & $8.9\times10^{5}$ & 1.97 & 0.326  \\
\hline\noalign{\smallskip}
\end{tabular}
\label{table_effA}
\end{table}

\begin{figure}[!ht]
  \centering
  \includegraphics[width=0.55\textwidth]{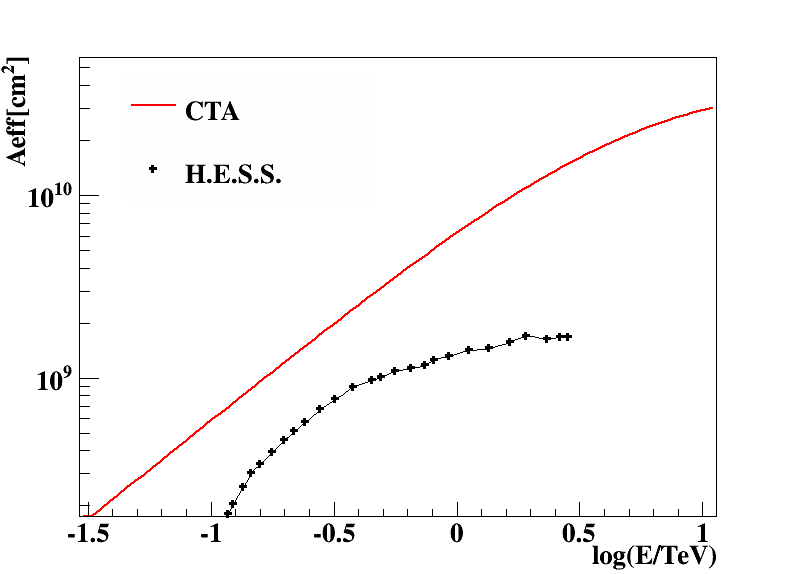}
  \caption{Effective area of CTA-South along with HESS effective area.}
  \label{effective_area}
\end{figure}

Figure \ref{effective_area} shows the considered effective area as a function of energy for a possible CTA-south array and for the HESS telescope array \citep{HESS2004sf2a.conf..407V}.
It can be seen from the figure that the effective area for CTA increases significantly compared to the one corresponding to HESS, specially for energies above 1 TeV. Such an improvement in the collection area also implies improving the telescope sensitivity. 

\begin{figure}[!ht]
  \centering
  \includegraphics[width=0.55\textwidth]{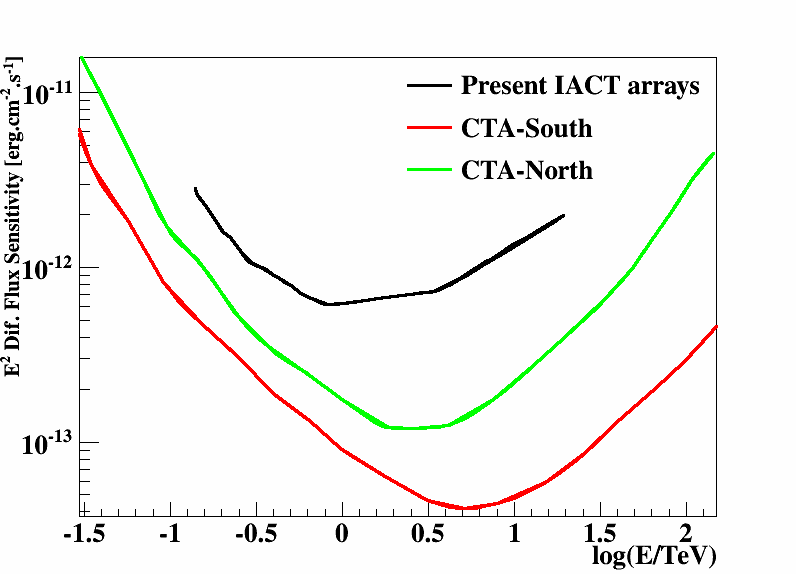}
  \caption{Sensitivity model for CTA-South (red line) and CTA-North (green line) taken from \cite{CTA_Concept} and sensitivity for a present telescope array (black). All curves correspond to a 50 hours observation time.}
  \label{sensitivity}
\end{figure}

Sensitivity is defined as the minimum flux of gamma rays required for a statistically significant detection in a given time. This is of course energy dependent and, in the case of current IACT, the energy boundaries for this flux lie between a few hundred GeV and 50 TeV. With the inclusion of four 23 m diameter Large Sized Telescopes (LST), CTA-south will lower the energy threshold down to $\sim 30$ GeV (see Figure \ref{sensitivity}), making it the first ground based telescope to significantly detect photons with energies in this range. 
An improved sensitivity and a lower energy threshold may increase chances of detecting an extended component, being lower energy photons the most affected, in principle, by the IGMF. However, as we will see in the next section, lowering the energy threshold is not necessarily always beneficial as there are other factors to be considered.

\subsection{Angular resolution and PSF}
                                                                                                               
The telescope angular resolution is probably the most relevant instrumental property for this study, as it is based on the discrimination between point like and non-point like angular distributions. Here we consider CTA angular resolution ($\sigma_{PSF}$) proposed in \cite{2016APh....80...22A}, which is described by the following expression,
\begin{equation}
\sigma_{PSF}\left(x\right)=\alpha\left[1+\exp\left( {-\frac{x}{\beta}}\right) \right], 
\end{equation}
where parameter $\alpha$ represents the achievable best resolution and parameter $\beta$ describes how fast the angular resolution changes with energy \citep{2016APh....80...22A}. The best fit parameters for the southern and northern sites are shown in Table \ref{table_ang}.
\begin{table}[!h]
\centering
\caption{Parameters of the fitted angular resolution for both sites.}

\begin{tabular}{cccc}
\hline\hline\noalign{\smallskip}
\!\! Site & \!\!\!\! $\alpha$ $[^\circ]$  & \!\!\!\! $\beta$  \\

\hline\noalign{\smallskip}
\!\! South  &  0.271  & 0.790 \\
\!\! North  &  0.291  & 0.763 \\
\hline\noalign{\smallskip}
\end{tabular}
\label{table_ang}
\end{table}
Given the IACT techniques for determining the gamma ray direction, higher energy gamma rays have more accurate direction determination than lower energy gamma rays. For this reason, the considered energy threshold for the analysis is critical, since it will ultimately determine the angular resolution of the instrument.

\begin{figure}[!ht]
  \centering
  \includegraphics[width=0.55\textwidth]{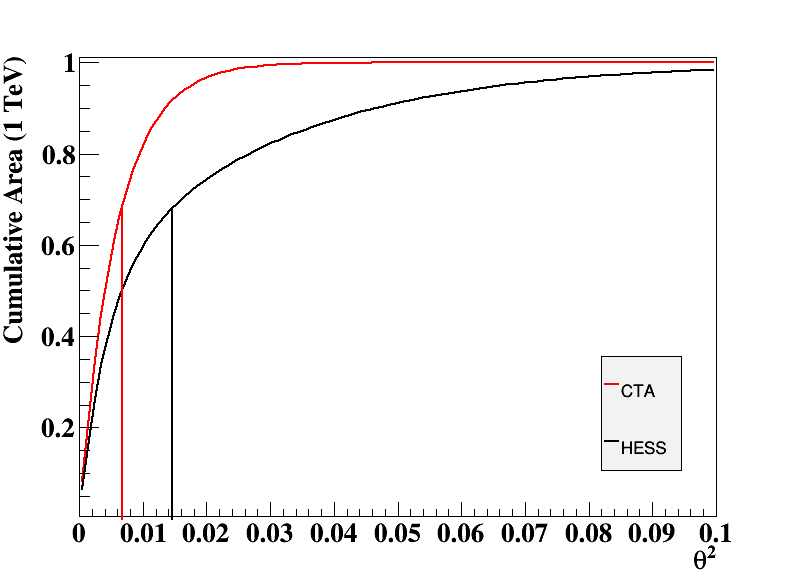}
  \caption{Normalized PSF cumulative distributions for CTA-south and HESS. Vertical lines represent the values of $\theta^2$ that encloses 68\% of the PSF area in each case.}
  \label{cumulative}
\end{figure}

The PSF, describes the angular response of the instrument to a point source, namely, the distribution of the angle ($\theta$) formed by the reconstructed photon arrival direction and the direction of the source. This function depends on the angular resolution which varies with the energy of the photon. The shape of this function, $f_{psf}$, can be described by a simple Gaussian function \citep{2016APh....80...22A},
\begin{equation}
f_{psf}\left(\theta^2\right)=\exp\left(-\frac{\theta^2}{2\sigma_{psf}^2}\right).
\label{PSF}
\end{equation} 
In this way, $\sigma_{psf}$ determines the 68\% containment radius of events ($\sigma_{68\%}^2=-2 \ln(0.32)\ \sigma_{psf}^2$) and that defines the angular resolution of the instrument considered for this study. Figure \ref{cumulative} shows the normalized PSF cumulative area distributions for CTA-south \citep{2016APh....80...22A} and HESS \citep{2006A&A...457..899A} at 1 TeV. Vertical lines represent the value of $\theta^2$ that encloses 68\% area of each PSF function; for this energy, CTA angular resolution drops down to below half the angular resolution achieved with the HESS telescope.

\subsection{Background rate}

Cosmic rays constitute the main background for ground based Cherenkov telescopes. This background is isotropic and has an approximately flat distribution in $\theta^2$. In this way, the background flux will unavoidably mix with the potential extended component of the source and the ability to distinguish this component will be directly affected by the background level. 
Present selection techniques allows to discriminate between gamma-ray and cosmic-ray events quite efficiently, however, a fraction of these events cannot be distinguished and are detected as gamma rays. The reason for this is that these selection techniques are usually based on the differences between the images of showers produced by cosmic and gamma rays, and these differences are not always clear, specially as the energy of the primary particle drops and the image reconstruction gets less accurate.
The background rate of the instrument is the number of cosmic-ray events that are systematically mistaken with gamma-ray events per unit of time. Having a good estimate of this rate allows to extract the right amount of background flux from the desired signal. For this study we use the background rate per unit of solid angle proposed in \cite{2016APh....80...22A}, that is energy dependent and it is described by the following expression,
\begin{center}
\begin{equation}
\begin{split}
B_{r}(x)=A_{1} \exp\left(-\frac{\left(x-\mu_{1}\right)^2}{2\sigma_{1}^2}\right)\\+A_{2} \exp\left(-\frac{\left(x-\mu_{2}\right)^2}{2\sigma_{2}^2}\right)+C
\end{split}
\label{BackEq}
\end{equation} 
\end{center}
The resulting parameters for the southern and northern site are displayed in Table \ref{table_back}.
\begin{table*}[!t]
\centering
\caption{Parameters for the background rate function for the southern and northern site.}

%\begin{tabular*}{\linewidth}{@{\extracolsep{\fill}}p{0.3\linewidth}p{0.3\linewidth}p{0.3\linewidth}@{}}
\begin{tabular}{cccccccc}
\hline\hline\noalign{\smallskip}
\!\! Site & \!\! A$_{1}$ [Hz/deg$^2$] & \!\! $\mu_{1}$  & \!\! $\sigma_{1}$ & \!\! A$_{2}$ [Hz/deg$^2$] & \!\! $\mu_{2}$ & \!\! $\sigma_{2}$ & \!\! C [Hz/deg$^2$] \\
\hline\noalign{\smallskip}
\!\! South  &  0.38 & -1.25 & 0.226 & 27.4                  & -3.90 & 0.998 & $3.78 \times 10^{-6}$\\
\!\! North  & 1.04  & -1.96 & 0.539 & $-2.83 \times 10^{4}$ & -10.4 & 0.114 & $1.93 \times 10^{-9}$\\
\hline\noalign{\smallskip}
\end{tabular}
\label{table_back}
\end{table*}
The background rate decreases with energy as expected. The overall background level is obtained by integrating this function between a chosen low energy threshold and 30 TeV. It should be noted that lowering the energy threshold means increasing the background level, this is an important fact to be considered when choosing the threshold for this and other analysis as well.

\section{Source Selection}

\begin{figure}[!ht]
  \centering
  \includegraphics[width=0.5\textwidth]{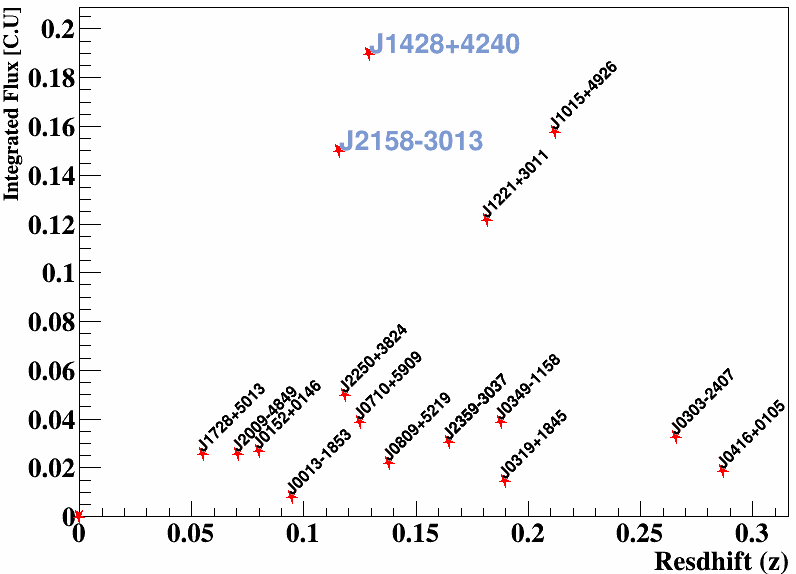}
  \caption{Integrated flux vs. redshift of a sample of 17 detected HBLs. Selected sources are highlighted in light blue.}
  \label{fluxvsz}
\end{figure}

There is a number of factors that make a particular source more likely to present a significant extended flux component than others. Distance/redshift is a crucial property: if the source is too close, the cascade process may not have enough distance to develop an appreciable broadening. If the source is too distant, the overall flux may be too low and the effect, although existent, will be too dimmed to be detected. Intrinsic spectral index is also a factor that could be determinant, since VHE gamma rays will produce a more energetic and richer cascades than lower energy photons, ``hard'' spectrum sources are more likely to present an appreciable extended flux component.
Blazars are by far the brightest extragalactic TeV sources, HBL subtype being the most numerous among the detected TeV sources. There are about 50 detected HBLs in the TeVCat catalog \citep{2008ICRC....3.1341W}, presenting a variety  of redshifts and fluxes. Figure \ref{fluxvsz} shows a sample of 17 HBLs with good redshift and flux determination. Flux and redshift data were taken from \cite{2008ICRC....3.1341W} and \cite{2016SPIE.9913E..1YC}.
For this study we considered two different blazars: PKS2155-304 visible from the southern hemisphere and H1426+428 visible from the northern hemisphere, which combine, in principle, good properties for developing an extended component. The corresponding TeV names for these sources as shown in Figure \ref{fluxvsz} are TeV-J2158-3013 and TeV-J1428+4240. Table \ref{source_param} shows a summary of the relevant parameters considered in this study for the selected sources.
\begin{table*}[!t]
\centering
\caption{Considered sources and their relevant parameters. $^\dagger$CU = $1.27 \times 10^{-11}$ cm$^{-2}$ s$^{-1}$ obtained by integrating Crab differential spectrum between 0.3 and 30 TeV.}

\begin{tabular}{cccc}
\hline\hline\noalign{\smallskip}
\!\! Source & \!\!\!\! Redshift & \!\!\!\! Spectral Index $\Gamma$ & \!\!\!\! Integrated Flux [cm$^{-2}$s$^{-1}$] - CU$^\dagger$  \\

\hline\noalign{\smallskip}
\!\! PKS 2155-304  &  0.11 & 1.83 & $1.9 \times 10^{-11}$ - 0.15 \\
\!\! H 1428+428    & 0.129 & 1.57 & $2.53x10^{-11}$ - 0.2 \\
\hline\noalign{\smallskip}
\end{tabular}
\label{source_param}
\end{table*}

Due to gamma-ray absorption and the cascade process taking place in the intergalactic medium, the observed spectral index is expected to be higher than the intrinsic one. The absorption process depends on the EBL properties and, although there are models that could be used to estimate the source intrinsic spectrum, EBL properties are rather uncertain. On the other hand, photons in the GeV range suffer almost no absorption and the observed spectral index should not differ too much with the intrinsic one at these energies. Following this argument, we consider the TeV part of the intrinsic spectra of the sources to be a prolongation of the GeV part. Therefore, the intrinsic spectrum should be well described by a power law with the observed GeV spectral index reported in the Fermi LAT 4-Year Point Source Catalog \citep{2015yCat..22180023A}. An exponential cutoff was set in 10 TeV to ensure a sufficient amount of VHE gamma rays \citep{2009IJMPD..18..911E}.
\begin{center}
\begin{equation}
\frac{dN}{dE}(E) = N_0\ E^{-\Gamma} \exp\left( -E/10\textrm{ TeV} \right),
\label{intrinsic}
\end{equation}
\end{center}
where $N_0$ is a normalization constant.

Flux variability is also an important factor to be considered. Blazars are specially known for their flux variability in the GeV-TeV range, presenting periods of high emission better known as flares. During these periods, flux can rapidly increase up to several orders of magnitude respect to the typical or quiescent flux of the source, and normal spectral properties of the source change, usually hardening the spectral index. 
As a result, a possible broadening effect in the emission can be outshined by the direct emission coming from the source. For this reason, the data sets used to study magnetic broadening are usually discriminated in those coming from high activity and low activity periods. In particular, the source PKS2155-304 has reported a relatively high variability index \citep{2015yCat..22180023A}, so the spectral properties considered in this study were taken from observations of the source during a quiescent state \citep{2010A&A...520A..83H}. Possible effects from past flares could, in principle, be present in the quiescent spectrum as well; these effects are commonly referred in the literature as pair echoes \citep{2011MNRAS.410.2741T}. Although pair echoes are not contemplated in this study, their possible effects would at least positively contribute to the broadening, so our considered scenarios should be pessimistic regarding this point.

\section{Simulations}
\label{SimulationsSection}

Intergalactic electromagnetic cascades are simulated using the Monte Carlo-based code Elmag \citep{Elmag2012CoPhC.183.1036K} version 2.03. The simulation injects photons (or leptons) into the intergalactic space at a chosen redshift, and takes into account pair production interactions with the EBL and the CMB, inverse Compton interactions of the resulting pairs and their synchrotron losses, and deflections in the IGMF. Input settings include an EBL model, IGMF coherence length and intensity, and the source injection spectrum. The simulation generates the spectral and angular distributions of photons when they arrive at Earth. The method presented in this work is strongly based on the simulated direction and energy of the arriving photons, so a few comments on how these are obtained and treated are necessary.
The simulation assumes a small deflection angle approximation, for strong IGMF intensities ($B>1\times10^{-14}$ G) where deflections start to be considerably large, specially for lower energetic particles, the program emulates an isotropic emission scenario by randomly assigning a direction to the scattered photons. For isotropically emitted photons, only those whose direction lies within the jet cone are considered, those scattered with bigger angles than the jet angle are dismissed. As a result, strong IGMF scenarios tend to show spectral energy distributions (SED) with a less populated low energy part. 

For this study we simulate 10 samples of $5\times10^5$ photons which are injected by the source following a power law intrinsic spectrum, as described by Eq.~(\ref{intrinsic}). Each sample is simulated with a different random initialization number to account for fluctuations coming from the MC process. The samples are then stacked together to improve statistics as much as possible. We assume the EBL model proposed in \cite{Dominguez} and an IGMF coherence length of 1 Mpc. A thinning factor of 0.3 is applied to speed up calculations. 
All simulations were done on an IGMF intensity grid within the range log(B/G)=-19 to log(B/G)=-14, including a null field ($B=0$ G) which is used as a non-existent IGMF scenario. Stronger fields ($>10^{-14}$G) were not considered in order to avoid conflict with the small angle approximation. Alternatively, we consider a different EBL scenario described by \cite{Gilmore} to study possible effects introduced by differences in the absorption process. The \cite{Gilmore} model is chosen because it presents considerable differences in the spectral shape in relation to the \cite{Dominguez} model, namely, a lower radiation density in the far-IR region. Finally, for the source PKS2155-304, two alternative spectral indexes are explored to study the effect of spectral differences in the output angular distributions. Sources that present a harder spectrum inject more high energy photons into the IG medium and are expected to produce a higher cascade component than soft sources. Two values are chosen, corresponding to a variation of 35$\%$ up and down from the spectral index value reported in the Fermi LAT 4-Year Point Source Catalog, i.e. $\Gamma=1.35$ and $\Gamma=2.30$.

\section{Method}

\begin{figure}[!ht]
  \centering
  \includegraphics[width=0.5\textwidth]{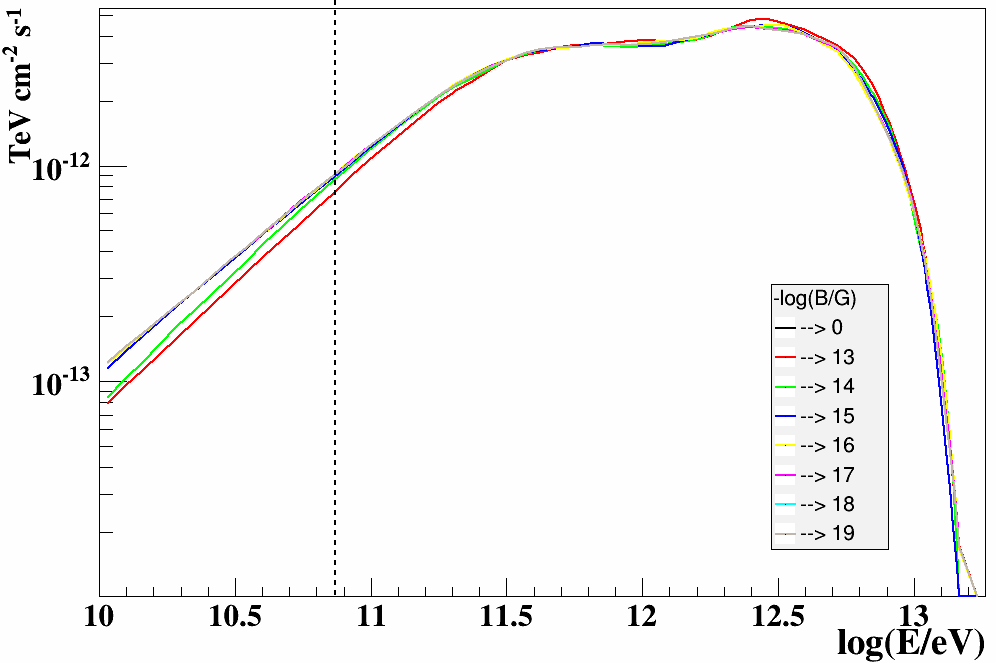}
  \caption{SED for different IGMF intensities considering the EBL model proposed in \cite{Dominguez}. The dashed vertical line represents the lower energy threshold of 75 GeV considered in this study.}
  \label{SED_Igmf}
\end{figure}

As mentioned before, there are several possible ways for searching IGMF imprints. The spectral energy distribution would in principle contain IGMF traces, particularly in the lower energy range, where the direction of the last scattering photons is determined by the electro-positron pairs that are more sensitive to the magnetic field strength. Figure \ref{SED_Igmf} shows spectral energy distributions for PKS2155-304 simulated for different IGMF intensities.
Spectra are well distinguished from each other in the lower energy range, specially for relatively high IGMF intensity scenarios. As the intensity weakens and the energy increases, spectra become indistinguishable from each other. In fact, above energies rounding 75 GeV, represented by the vertical dashed line in Figure \ref{SED_Igmf}, SEDs are practically indistinguishable for any IGMF scenario with $B\leq 10^{-14}$. On the other hand, the angular distribution of photons is also expected to contain IGMF traces and could potentially lead to constrains or even a detection. Any significant difference between a source angular distribution and the one corresponding to a point source, i.e. the telescope PSF, would indicate the existence of an extended contribution. Our strategy consists in comparing simulated distributions of non-zero IGMF scenarios with a point like distribution constructed from a zero IGMF simulation; we test the power of a CTA-like instrument to distinguish the angular distributions of those two simulations. More specifically we test, for a given IGMF scenario, the probability of rejecting the null hypothesis in which non-zero and null IGMF $\theta^2$ distributions come from the same parent distribution.

\subsection{PSF convolution}
\label{psf}

To obtain a realistic representation of the observed angular distributions, the output of the simulated events are convolved with the telescope PSF. The convolution process is done by getting the energy of each simulated event and randomly sampling Eq.~(\ref{PSF}) evaluated in that energy. The resulting value of $\theta^2_{psf}$ is then used to obtain the reconstructed arrival direction of the photon, $\theta^2$, by using following expression,
\begin{equation}
\theta^2=\theta^2_{sim}+\theta^2_{psf}+2\ \theta_{sim}\ \theta_{psf} \cos\phi
\end{equation}
where $\theta^2_{sim}$ is obtained from the Elmag simulation and $\phi$ (azimuth angle) is uniformly sampled from the interval $(0,2\pi]$ to account for the symmetry under azimuthal rotation. Each event is also weighted considering the telescope effective area corresponding to the event energy. Finally, an overall factor is applied to normalize the total number of events according to the source integrated flux (see table \ref{source_param}), which is obtained by integrating the observed spectrum of each source, and a given observation time. 

\subsection{Fluctuations and background subtraction}

Fluctuations and background events are added to the convolved simulated distribution corresponding to an IGMF of $B=0$. To achieve this, each bin content $\mu_i$ of the $\theta^2$ histogram is used to sample a Poisson distribution, which is given by,
\begin{equation}
P(n;\mu)=\frac{\mu^n}{n!}e^{-\mu}
\label{Poiss}
\end{equation}
where $\mu=\mu_i$ for the $i$-th bin. The sampled value, $n_i$, is then used to construct a new fluctuated distribution $H_{fluct}\left(\theta^2\right)$ where each bin content is a different Poisson-fluctuated value of the original bin content. Additionally, the average number of background events is obtained by integrating the background rate given by Eq.~(\ref{BackEq}) between the corresponding lower energy threshold and 30 TeV,
\begin{center}
\begin{equation}
\mu_{b}=\pi ~T_{obs}~ \Delta\theta~\int_{E_{th}}^{30\ \textrm{TeV}} B_r(E)~dE
\end{equation}
\end{center}
where $B_e$ is the background rate, T$_{obs}$ is the observation time, and $\Delta\theta$ the bin width of the histogram under consideration. The resulting value  $\mu_{b}$  is then used to sample a Poisson distribution (Eq.~(\ref{Poiss}) with $\mu=\mu_b$). The sampled background events are added bin-wise to the $H_{fluct}\left(\theta^2\right)$ distribution to obtain a raw observed distribution $H_{raw}\left(\theta^2\right)$. Finally the background is subtracted following the Wobble method \citep{Wobble2011arXiv1111.0121F}, where 3 different patches corresponding to the off-source region of the sky are simultaneously observed along the on-source region, and then used for background subtraction. This procedure was emulated by sampling a Poisson distribution with mean value $\mu_W=3\ \mu_b$. The sampled value is then subtracted from each bin of the $H_{raw}\left(\theta^2\right)$ distribution to obtain the excess distribution $H_{excess}\left(\theta^2\right)$.

\subsection{Rejection}
\label{rejec_sect}

Non null magnetic field models, obtained in the same way as the PSF (see Sec.~\ref{psf}) but in this case without including neither Poisson fluctuations nor background, are then fitted to the excess distribution ($H_{excess}$). The fitting parameter is a normalization constant which is obtained by minimizing the $\chi^2$. The analytic expression for the parameter is given by,   
\begin{equation}
A=\frac{\sum_{i}\left(H_{excess,\, i}~H_{model,\, i}\right)/\sigma_i^2}{\sum_{i} H^2_{model,\, i}/\sigma_i^2},
\end{equation}
where $H_{excess,\, i}$ and $H_{model,\, i}$ are the contents of the $i$-th bin for the excess and model distributions respectively, and $\sigma_i$ represents the standard deviation of the $i$-th bin of the the excess distribution. By using the Possonian character of all fluctuation the variance of $i$-th bin content can be estimated as,
\begin{equation}
\sigma_i^2=H_{raw,\, i}+\mu_{b,\, i}+\frac{\mu_{W,\, i}}{3}.
\end{equation} 
Finally, a $\chi^2$ test is applied to obtain the p-value of the fit. This process is repeated 1000 times for every IGMF intensity, and each time with a different fluctuated excess distribution to emulate different possible observations. Our null hypothesis is to consider the excess PSF distribution ($B=0$) and the simulated model (non-null magnetic field) as coming from the same parent distribution. The probability of rejecting this hypothesis with a 99$\%$ confidence level is then given by the number of cases whose p-value lies below 0.01, over the total number of cases considered.

\section{Results and discussion}
 
\begin{figure}[!ht]
  \centering
  \includegraphics[width=0.5\textwidth]{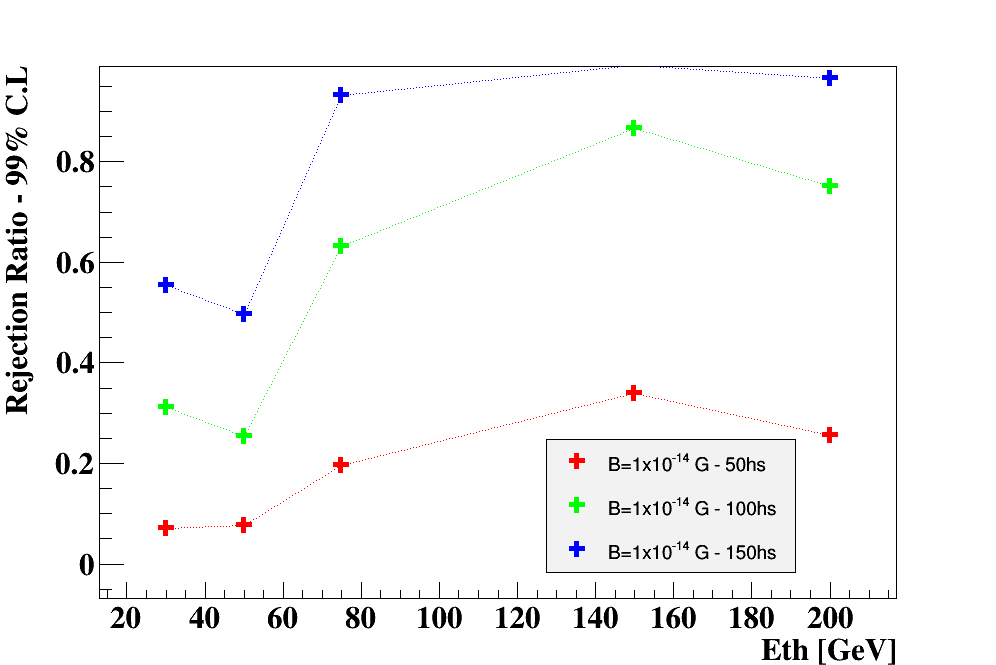}
  \caption{Rejection probabilities for PKS 2155-304 (CTA-south) for different low energy thresholds. The results correspond to the IGMF scenario with intensity $B=10^{-14}$ G.}
  \label{rejection_PKS_Eth}
\end{figure}

A preliminary study was performed to determine the optimal low energy threshold to be considered in the analysis. In previous sections we discussed how lowering the energy threshold will not necessarily enhance the cascade effect in the overall distributions. Many of the telescope performance parameters, such as background level and angular resolution, will be inevitably affected negatively as the threshold energy decreases. To estimate the impact of the energy threshold we considered five different low threshold values: 30 GeV, 50 GeV, 75 GeV, 150 GeV, and 200 GeV and calculated the rejection ratio for an IGMF intensity of $B=10^{-14}$ G and for three different values of the observation time, 50, 100, and 150 hours. Figure \ref{rejection_PKS_Eth} shows the resulting rejection probabilities for each case.
It is clear from this example that neither the lowest nor the highest energy thresholds present the best probabilities of detecting a possible broadening. It might be tempting to say that 150 GeV is the optimal energy threshold, however, the results shown represent only the behavior of one particular source and one particular IGMF intensity. The rejection ratio trend with energy threshold is likely to depend on the source characteristics and even on the IGMF intensity. For this reasons a more conservative 75 GeV lower energy threshold is adopted throughout the rest of the analysis.

\begin{figure}[!ht]
  \centering
  \includegraphics[width=0.5\textwidth]{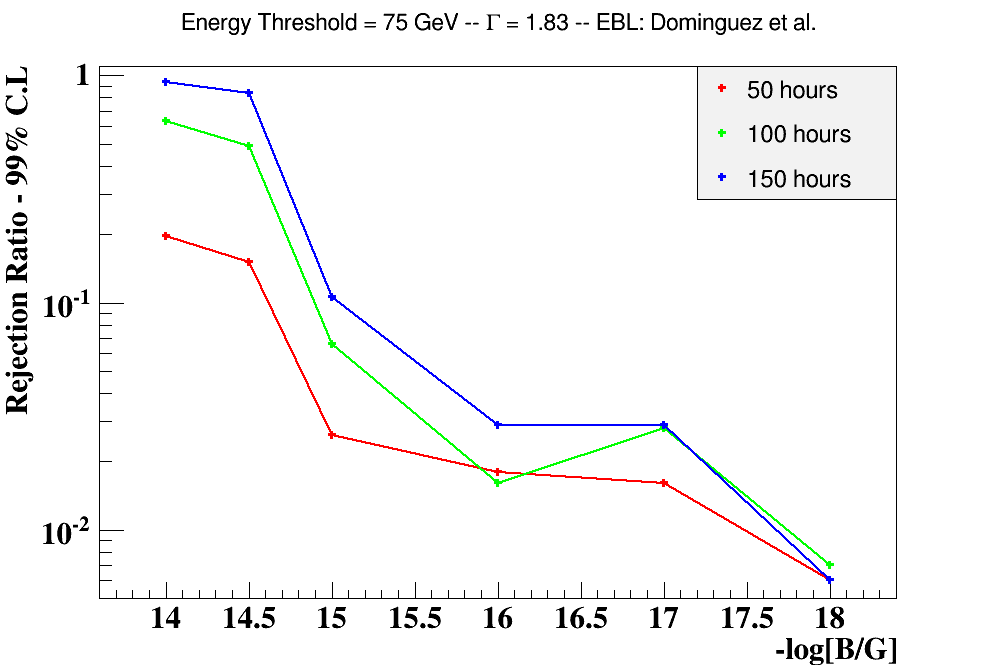}
  \caption{Rejection probabilities for PKS 2155-304 (CTA-south) for different observation time windows.}
  \label{rejection_PKS_t}
\end{figure}

\begin{figure}[!ht]
  \centering
  \includegraphics[width=0.5\textwidth]{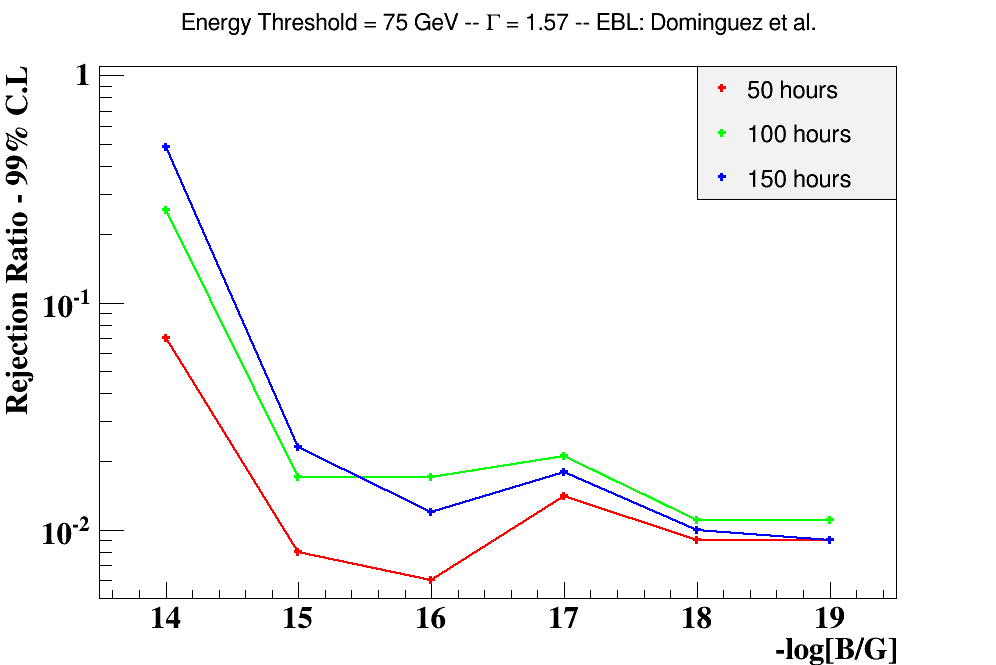}
  \caption{Rejection probabilities for H1426+428 (CTA-north) for different observation time windows.}
  \label{rejection_H1426}
\end{figure}

Figures \ref{rejection_PKS_t} and \ref{rejection_H1426} show the resulting rejection probabilities for PKS2155-304 and H1426+428 respectively, as a function of the IGMF intensity and for a low energy threshold of 75 GeV and for 50, 100, and 150 hours of observation time.
%Each IGMF scenario is simulated 10 times using different random initial seeds to estimate the intrinsic fluctuations coming from the simulation. These fluctuations are then used to calculate the average value and the standard deviation of each probability point.  
In general, the probability is considerably higher IGMF intensities above $\sim$10$^{-14}$ G, where it increases with longer observation time windows. The rejection ratio then drops down for weaker magnetic fields in both the southern and northern sites, however, in the latter the ratios are smaller and the drop is more abrupt, probably because of the differences in the instrument performance. The drop in the rejection ratio for weak magnetic fields responds to the fact that in these scenarios, the cascade component is too weak and/or not broadened enough to be differentiated from the PSF distribution and its fluctuations. 

\begin{figure}[!ht]
  \centering
  \includegraphics[width=0.5\textwidth]{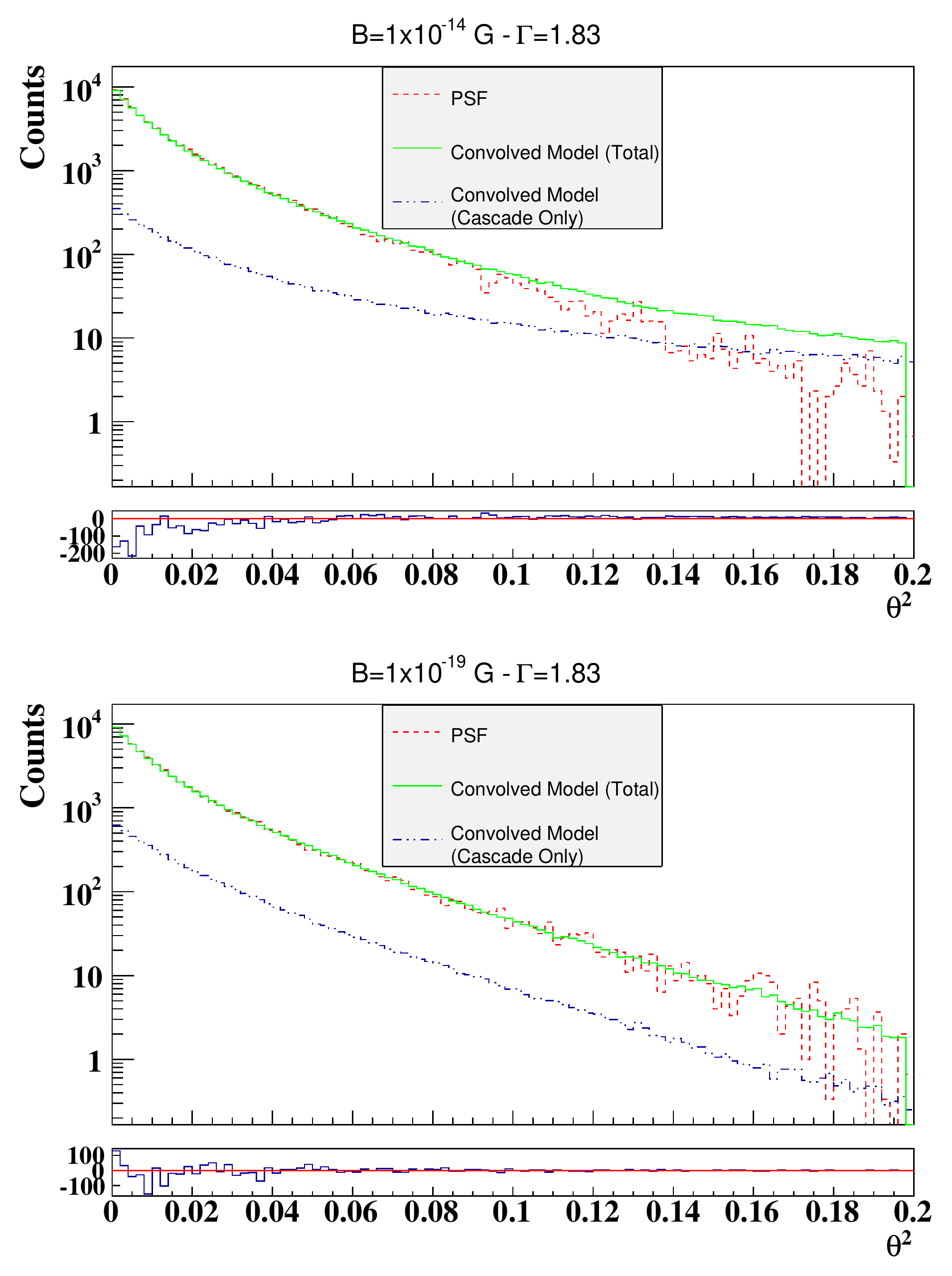}
  \caption{Fluctuated PSF $\theta^2$ distribution fitted with the $\theta^2$ distributions corresponding to $B=10^{-14}$ G (left) and $B=10^{-19}$ G (right). Also shown are the $\theta^2$ distributions corresponding to the cascade photons. Lower pads show the bin content difference between the convolved model $\theta^2$ distribution and the PSF distribution in each case.}
  \label{ThetaCompared}
\end{figure}

We have discussed how stronger IGMF scenarios lead to broader angular dispersion of the arriving photons, which can be directly appreciated in the $\theta^2$ distributions of the arriving photons. Figure \ref{ThetaCompared} shows the fits of PKS2155-304 PSF, for two particular samples, with the $\theta^2$ distributions (after the PSF convolution) corresponding to two IGMF intensities, a strong field ($B = 10^{-14}$ G) and a weak field ($B = 10^{-19}$ G). Also shown is the $\theta^2$ distributions of the cascade photons. 
The convolved model $\theta^2$ distribution is clearly broader than the PSF distribution in the case of strong magnetic field, whereas for the weak magnetic field scenario these two distributions are almost indistinguishable. The amount of photons within the cascade distribution and its shape, will ultimately determine whether or not the broadening is appreciable. 

\begin{figure}[!ht]
  \centering
  \includegraphics[width=0.5\textwidth]{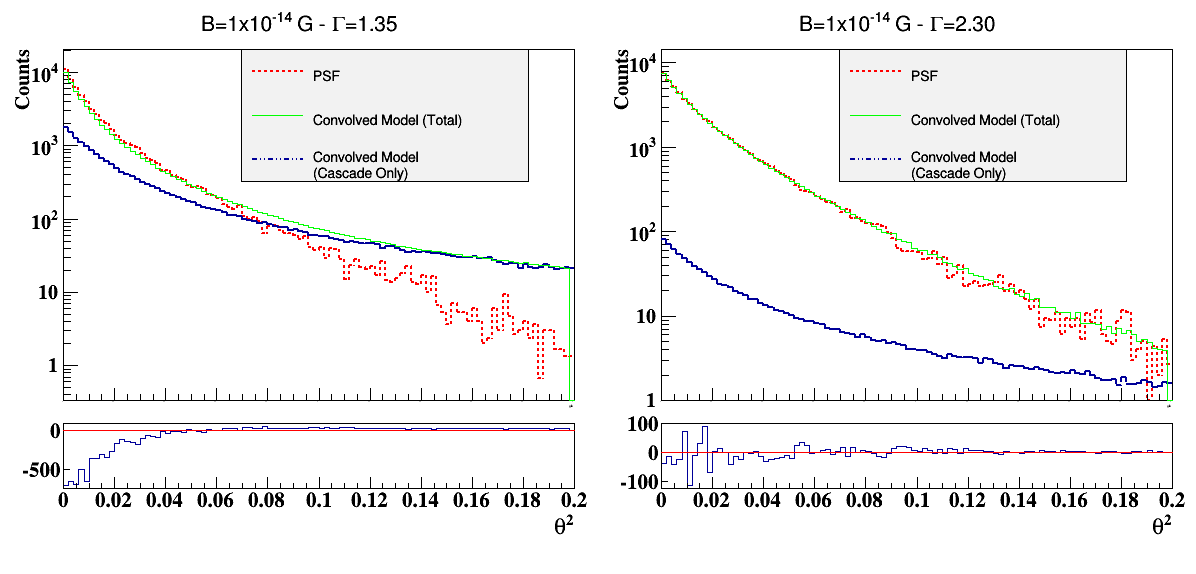}
  \caption{Fluctuated PSF $\theta^2$ distribution fitted with the $\theta^2$ distributions corresponding to $B=10^{-14}$ G for spectral indexes $\Gamma=1.35$ (left) and $\Gamma=2.30$ (right). Also shown are the $\theta^2$ distributions corresponding to the cascade photons. Lower pads show the bin content difference between the convolved model $\theta^2$ distribution and the PSF distribution in each case.}
  \label{ThetaCompared_alt}
\end{figure}

\begin{figure}[!ht]
  \centering
  \includegraphics[width=0.5\textwidth]{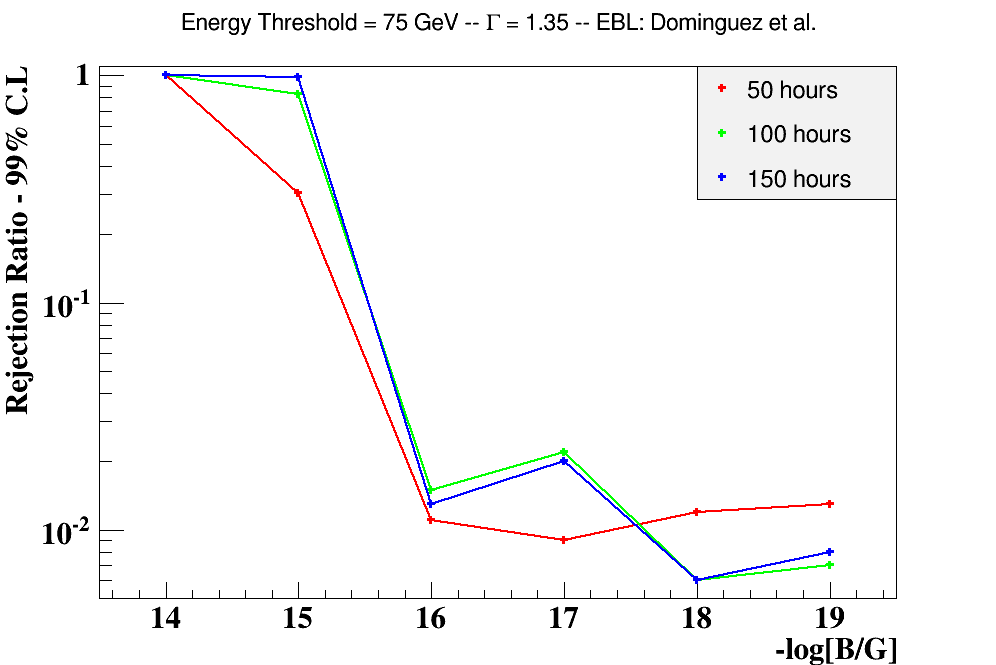}
  \includegraphics[width=0.5\textwidth]{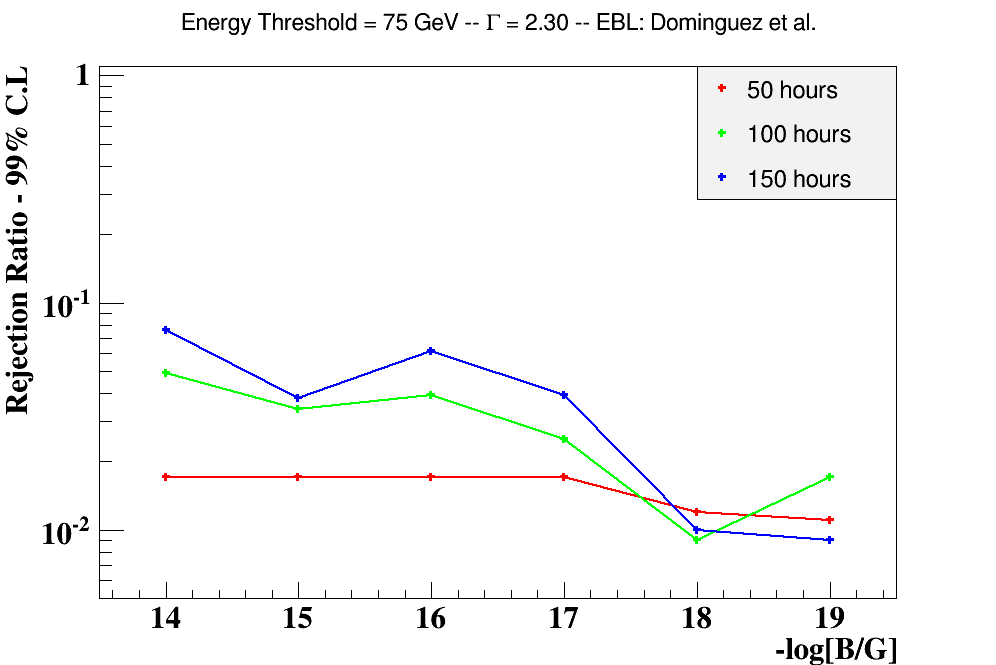}
  \caption{Rejection ratios for PKS 2155-304 (CTA-south) with alternative spectral indexes of $\Gamma=1.35$ (top) and $\Gamma=2.30$ (bottom).}
  \label{Rejec_alt}
\end{figure}

The IGMF intensity is not the only factor that affects the shape and broadening of the cascade distribution. A less direct IGMF effect imprinted in the $\theta^2$ distribution comes from the energy spectrum of the photons emitted from the source. As discussed in Sec.\ref{SimulationsSection}, hard spectrum sources are expected to inject a higher amount of high energy photons capable of generating secondary gamma rays and thus resulting in a relatively more important cascade component. Figure \ref{ThetaCompared_alt} shows the fits of PKS2155-304 PSF, for two particular samples, with the $\theta^2$ distributions (after the PSF convolution) corresponding to $B = 10^{-14}$ G and for two alternative spectral indexes of $\Gamma=1.35$ and $\Gamma=2.30$. Also shown is the $\theta^2$ distributions of the cascade photons.
The spectral index effect can be clearly appreciated in both the flux level and shape of the cascade distributions. The cascade component in the case of $\Gamma=1.35$ represents a significant part of the total flux, and the shape of its distribution clearly differs from the shape of the PSF. On the other hand, the $\theta^2$ distribution for $\Gamma=2.30$ has a much less significant cascade component and the overall distribution cannot be distinguished from the PSF and its fluctuations. Figure \ref{Rejec_alt} shows the rejection probabilities for these last two cases with alternative spectral indexes.
As expected, the hard spectrum case shows, in general, higher rejection ratios than the cases with softer spectral index. For the IGMF scenario with $B=10^{-14}$ G, the null hypothesis is rejected regardless of the observation time. The hard spectrum case also presents a smoother transition to weaker IGMF intensities, showing promising results for scenarios with intensities of $\sim$10$^{-15}$G. On the other hand, the softer spectral index shows small rejection ratios for all IGMF scenarios, reinforcing the idea that the IGMF studies are less promising when soft spectrum sources are considered. Regarding the energy threshold discussed previously, the hard spectrum case shows an improvement in the rejection ratios as the energy threshold decreases. This can be attributed to the fact that most of the secondary photons that constitute the cascade component, populate the lower energy range of the SED, and being the case with most significant cascade component, it is expected that a lower cut in energy will result in an enhancement of the cascade distribution.

\begin{figure}[!ht]
  \centering
  \includegraphics[width=0.5\textwidth]{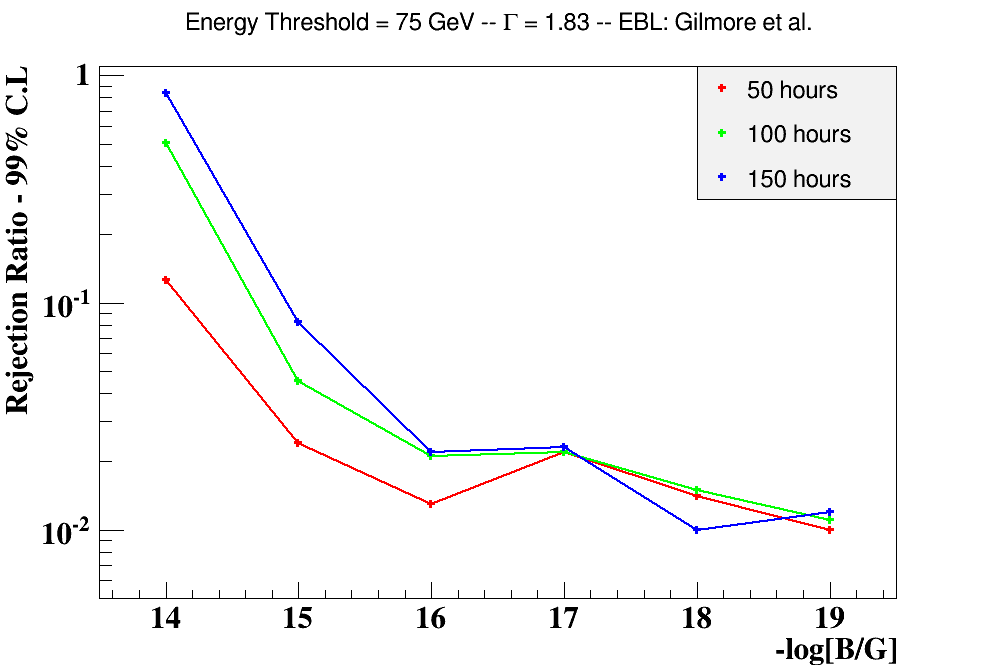}
  \caption{Rejection ratios for PKS 2155-304 (CTA-south) considering the EBL model porposed in \cite{Gilmore}.}
  \label{rejection_PKS_Gilmore}
\end{figure}

Figure \ref{rejection_PKS_Gilmore} shows the rejection ratios corresponding to PKS 2155-304 when considering the EBL model proposed in \cite{Gilmore}. Any difference in the absorption process that could be introduced by changing the EBL model are simply too small to be appreciated in the current $\theta^2$ distributions and the resulting rejection ratios are practically the same as the ones obtained with the \cite{Dominguez} EBL model.  

Alternative intrinsic source spectra for PKS 2155-304 with energy cut-offs in E$_{cut}$=5 TeV and E$_{cut}$=20 TeV were considered to estimate the impact of high energy photons in the broadening effect. Results show that in the case of a 5 TeV cut-off, rejection probabilities drop below 0.1 for all fields and observation times. This is consistent with the fact that, for this case, fewer high energy photons are being injected in the intergalactic medium, producing a less significant amount of cascade photons. On the other hand, a cut-off in 20 TeV shows no significant changes in relation to the 10 TeV cut-off spectrum, perhaps suggesting that, in spite of the higher cut-off, the amount of injected photons is not enough to produce detectable improvements in the rejection ratios. If this is the case, a plateau in the rejection ratios would be expected even for higher energy cut-offs.

\begin{figure}[!ht]
  \centering
  \includegraphics[width=0.5\textwidth]{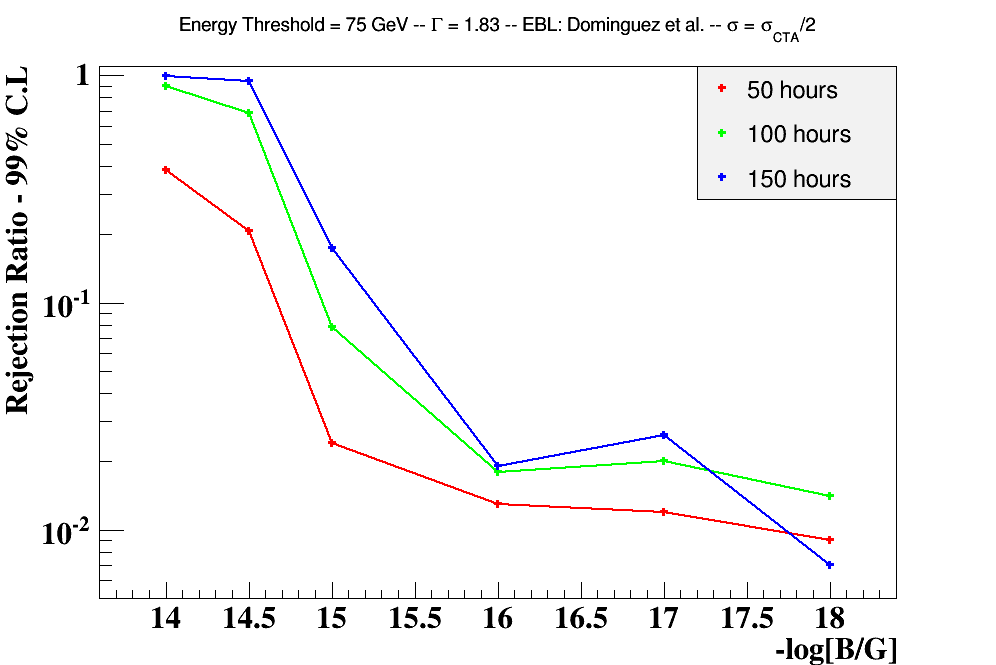}
  \caption{Rejection ratios for PKS 2155-304 (CTA-south) considering a narrower PSF with $\sigma_{psf}/2$.}
  \label{rejection_PKS_HalfSigma}
\end{figure}

To estimate the limitations due to the instrument performance, the rejection probabilities were also calculated considering a narrower PSF. This was achieved by halving the 68$\%$ containment radius of the PSF, $\sigma_{psf}$ in Eq. \ref{PSF}. Results for the improved instrument are shown in figure \ref{rejection_PKS_HalfSigma}.

Although probabilities for this case are slightly higher, pushing the detection limit towards weaker IGMF intensities, the values are still too low to claim any detection around $\sim$10$^{-15}$G. However, it is reasonable to expect higher values as the angular resolution of the instrument improves.  

\section{Conclusions}

The possible magnetic broadening effect in the angular distribution of gamma rays coming from distant blazars constitutes an alternative method to study and constrain the IGMF. The aim of this study is to try to asses quantitatively the detection of possible magnetic broadening with next generation Cherenkov telescopes, given a realistic set of observations and instrumental response. The detection will ultimately rely on which source or sources are chosen for the study and what method is used for discriminating the cascade component. 
For the analysed source PKS 2155-304 corresponding to the southern site, results show that CTA should be able to detect IGMF with intensities stronger than 10$^{-14.5}$G within an observation time of $\sim$ 100 hours. The source H1426+428 corresponding to the northern site shows a similar trend, although in this case rejection ratio values are lower, probably due to the instrumental limitations. The obtained results also give us some valuable information on what factors are specially determinant. Source spectral index and flux seem to be key properties. The source PKS 2155-304 shows significant changes when the the spectral index is varied by 30\%, a reasonable amount given the uncertainties in this parameter. The soft version of its spectrum shows low rejection probabilities for all scenarios, whereas the hard one increases them and pushes the detection range down to IGMF intensities of $\sim10^{-15}$G. Spectral index and flux level should be prioritizing properties when looking for suitable source candidates. The instrument performance is also determinant. There is a clear difference in the results coming from the 50, 100, and 150 hour observation times. In this sense, the effective area of the instrument will ultimately determine whether is practical or not to expect positive results from typical observation times. Background fluctuations are also related to the observation time, and they are crucial when it comes to discriminating the extended component. Any improvement in the treatment of these fluctuations will certainly have a positive impact in the detection of the broadening. Finally, the method also relies on the angular resolution of the instrument, and although our method shows slightly better results for a PSF with half the width, it is expected that future experiments will increase significantly their chances as they improve their angular resolution.
\newline

This work has been partially supported with a grant from Agencia Nacional de Promoci\'on Cient\'ifica y Tecnol\'ogica, Argentina. The author M.F.A. developed this work with a Doctoral fellowship from CONICET, Argentina. The authors A.D.S. and A.C.R. are members of the Carrera del Investigador Cienti\'ifico of CONICET, Argentina. We thank the anonymous referee for a careful reading and useful questions and suggestions that improved the presentation of this paper.

\bibliographystyle{aasjournal}
\bibliography{biblio_fernandez}

\begin{thebibliography}{}
\expandafter\ifx\csname natexlab\endcsname\relax\def\natexlab#1{#1}\fi
\providecommand{\url}[1]{\href{#1}{#1}}
\providecommand{\dodoi}[1]{doi:~\href{http://doi.org/#1}{\nolinkurl{#1}}}
\providecommand{\doeprint}[1]{\href{http://ascl.net/#1}{\nolinkurl{http://ascl.net/#1}}}
\providecommand{\doarXiv}[1]{\href{https://arxiv.org/abs/#1}{\nolinkurl{https://arxiv.org/abs/#1}}}

\bibitem[{CTA(2017)}]{CTAperf}
 2017, {CTA} Performance Files,
  \url{https://portal.cta-observatory.org/CTA_Observatory/performance/SitePages/Home.aspx}

\bibitem[{{Abramowski} {et~al.}(2014){Abramowski}, {Aharonian}, {Ait Benkhali},
  {Akhperjanian}, {Ang{\"u}ner}, {Anton}, {Backes}, {Balenderan}, {Balzer}, \&
  et~al.}]{Abramow2014A&A...562A.145H}
{Abramowski}, A., {Aharonian}, F., {Ait Benkhali}, F., {et~al.} 2014, Astron.
  Astrophys., 562, A145, \dodoi{10.1051/0004-6361/201322510}

\bibitem[{{Acero} {et~al.}(2015){Acero}, {Ackermann}, \&
  {Ajello}}]{2015yCat..22180023A}
{Acero}, F., {Ackermann}, M., \& {Ajello}, M. 2015, VizieR Online Data Catalog,
  221

\bibitem[{{Acharya} {et~al.}(2013){Acharya}, {Actis}, {Aghajani}, {Agnetta},
  {Aguilar}, {Aharonian}, {Ajello}, {Akhperjanian}, {Alcubierre},
  {Aleksi{\'c}}, \& et~al.}]{CTA_Concept}
{Acharya}, B.~S., {Actis}, M., {Aghajani}, T., {et~al.} 2013, Astroparticle
  Physics, 43, 3, \dodoi{10.1016/j.astropartphys.2013.01.007}

\bibitem[{{Ackermann} {et~al.}(2018){Ackermann}, {Ajello}, \&
  {Baldini}}]{2018ApJS..237...32A}
{Ackermann}, M., {Ajello}, M., \& {Baldini}, L. 2018, \apjs, 237, 32,
  \dodoi{10.3847/1538-4365/aacdf7}

\bibitem[{{Aharonian} \& {Akhperjanian}(2006)}]{2006A&A...457..899A}
{Aharonian}, F., \& {Akhperjanian}, A. 2006, Astron. Astrophys., 457, 899,
  \dodoi{10.1051/0004-6361:20065351}

\bibitem[{{Aharonian} {et~al.}(2001){Aharonian}, {Akhperjanian}, {Barrio},
  {Bernl{\'o}hr}, {Bolz}, \& {B{\'o}rst}}]{2001A&A...366..746A}
{Aharonian}, F., {Akhperjanian}, A., {Barrio}, J., {et~al.} 2001, Astron.
  Astrophys., 366, 746, \dodoi{10.1051/0004-6361:20000481}

\bibitem[{{Aharonian} {et~al.}(1994){Aharonian}, {Coppi}, \&
  {Voelk}}]{1994ApJ...423L...5A}
{Aharonian}, F., {Coppi}, P., \& {Voelk}, H. 1994, Astrophys. J. Lett, 423, L5,
  \dodoi{10.1086/187222}

\bibitem[{{Aharonian} {et~al.}(1997){Aharonian}, {Hofmann}, {Konopelko}, \&
  {V{\"o}lk}}]{1997APh.....6..369A}
{Aharonian}, F.~A., {Hofmann}, W., {Konopelko}, A.~K., \& {V{\"o}lk}, H.~J.
  1997, Astroparticle Physics, 6, 369, \dodoi{10.1016/S0927-6505(96)00070-9}

\bibitem[{{Aharonian} {et~al.}(2010){Aharonian}, {Kelner}, \&
  {Prosekin}}]{Aharonian2010PhRvD..82d3002A}
{Aharonian}, F.~A., {Kelner}, S.~R., \& {Prosekin}, A.~Y. 2010, \prd, 82,
  043002, \dodoi{10.1103/PhysRevD.82.043002}

\bibitem[{{Ahlers}(2011)}]{Ahlers2011PhRvD..84f3006A}
{Ahlers}, M. 2011, \prd, 84, 063006, \dodoi{10.1103/PhysRevD.84.063006}

\bibitem[{{Aleksi{\'c}} {et~al.}(2010){Aleksi{\'c}}, {Antonelli}, {Antoranz},
  {Backes}, {Baixeras}, \& {Barrio}}]{2010A&A...524A..77A}
{Aleksi{\'c}}, J., {Antonelli}, L., {Antoranz}, P., {et~al.} 2010, Astron.
  Astrophys., 524, A77, \dodoi{10.1051/0004-6361/201014747}

\bibitem[{{Ambrogi} {et~al.}(2016){Ambrogi}, {De O{\~n}a Wilhelmi}, \&
  {Aharonian}}]{2016APh....80...22A}
{Ambrogi}, L., {De O{\~n}a Wilhelmi}, E., \& {Aharonian}, F. 2016,
  Astroparticle Physics, 80, 22, \dodoi{10.1016/j.astropartphys.2016.03.004}

\bibitem[{{Archambault} {et~al.}(2017){Archambault}, {Archer}, \&
  {Benbow}}]{2017ApJ...835..288A}
{Archambault}, S., {Archer}, A., \& {Benbow}, W. 2017, \apj, 835, 288,
  \dodoi{10.3847/1538-4357/835/2/288}

\bibitem[{{Arlen} {et~al.}(2014){Arlen}, {Vassilev}, {Weisgarber}, {Wakely}, \&
  {Yusef Shafi}}]{Arlen2014ApJ...796...18A}
{Arlen}, T.~C., {Vassilev}, V.~V., {Weisgarber}, T., {Wakely}, S.~P., \& {Yusef
  Shafi}, S. 2014, Astrophys. J., 796, 18, \dodoi{10.1088/0004-637X/796/1/18}

\bibitem[{{Bertone} {et~al.}(2006){Bertone}, {Vogt}, \&
  {En{\ss}lin}}]{2006MNRAS.370..319B}
{Bertone}, S., {Vogt}, C., \& {En{\ss}lin}, T. 2006, Mon. Not. R. Astron Soc,
  370, 319, \dodoi{10.1111/j.1365-2966.2006.10474.x}

\bibitem[{{Blasi} {et~al.}(1999){Blasi}, {Burles}, \&
  {Olinto}}]{1999ApJ...514L..79B}
{Blasi}, P., {Burles}, S., \& {Olinto}, A.~V. 1999, Astrophys. J. Lett, 514,
  L79, \dodoi{10.1086/311958}

\bibitem[{{Broderick} {et~al.}(2012){Broderick}, {Chang}, \&
  {Pfrommer}}]{2012ApJ...752...22B}
{Broderick}, A.~E., {Chang}, P., \& {Pfrommer}, C. 2012, \apj, 752, 22,
  \dodoi{10.1088/0004-637X/752/1/22}

\bibitem[{{Caprini} \& {Gabici}(2015)}]{Caprini2015PhRvD..91l3514C}
{Caprini}, C., \& {Gabici}, S. 2015, \prd, 91, 123514,
  \dodoi{10.1103/PhysRevD.91.123514}

\bibitem[{{Carosi} {et~al.}(2016){Carosi}, {Lucarelli}, {Antonelli}, \&
  {Giommi}}]{2016SPIE.9913E..1YC}
{Carosi}, A., {Lucarelli}, F., {Antonelli}, L.~A., \& {Giommi}, P. 2016, in
  Software and Cyberinfrastructure for Astronomy IV, Vol. 9913, 99131Y

\bibitem[{{Chen} {et~al.}(2015){Chen}, {Buckley}, \&
  {Ferrer}}]{Chen2015PhRvL.115u1103C}
{Chen}, W., {Buckley}, J.~H., \& {Ferrer}, F. 2015, Physical Review Letters,
  115, 211103, \dodoi{10.1103/PhysRevLett.115.211103}

\bibitem[{{Dom{\'{\i}}nguez} {et~al.}(2011){Dom{\'{\i}}nguez}, {Primack},
  {Rosario}, {Prada}, {Gilmore}, {Faber}, {Koo}, {Somerville},
  {P{\'e}rez-Torres}, {P{\'e}rez-Gonz{\'a}lez}, {Huang}, {Davis},
  {Guhathakurta}, {Barmby}, {Conselice}, {Lozano}, {Newman}, \&
  {Cooper}}]{Dominguez}
{Dom{\'{\i}}nguez}, A., {Primack}, J.~R., {Rosario}, D.~J., {et~al.} 2011, Mon.
  Not. R. Astron Soc, 410, 2556, \dodoi{10.1111/j.1365-2966.2010.17631.x}

\bibitem[{{Durrer} \& {Neronov}(2013)}]{2013A&ARv..21...62D}
{Durrer}, R., \& {Neronov}, A. 2013, Astron. Astrophys., 21, 62,
  \dodoi{10.1007/s00159-013-0062-7}

\bibitem[{{Eungwanichayapant} \& {Aharonian}(2009)}]{2009IJMPD..18..911E}
{Eungwanichayapant}, A., \& {Aharonian}, F. 2009, International Journal of
  Modern Physics D, 18, 911, \dodoi{10.1142/S0218271809014832}

\bibitem[{{Fallon}(2010)}]{2010tsra.confE.192F}
{Fallon}, L. 2010, in 25th Texas Symposium on Relativistic Astrophysics.

\bibitem[{{Fernandez Alonso}(2014)}]{2014arXiv1406.4764F}
{Fernandez Alonso}, M. 2014, BAAA 56.
\newblock \doarXiv{1406.4764}

\bibitem[{{Fernandez Alonso} {et~al.}(2015){Fernandez Alonso}, {Supanitsky}, \&
  {Rovero}}]{Fernandez2015ICRC...34..791F}
{Fernandez Alonso}, M., {Supanitsky}, A.~D., \& {Rovero}, A.~C. 2015, in
  International Cosmic Ray Conference, Vol.~34, 34th International Cosmic Ray
  Conference (ICRC2015), 791

\bibitem[{{Finke} {et~al.}(2015){Finke}, {Reyes}, {Georganopoulos}, {Reynolds},
  {Ajello}, {Fegan}, \& {McCann}}]{Finke2015ApJ...814...20F}
{Finke}, J.~D., {Reyes}, L.~C., {Georganopoulos}, M., {et~al.} 2015, Astrophys.
  J., 814, 20, \dodoi{10.1088/0004-637X/814/1/20}

\bibitem[{{Finnegan} \& {for the VERITAS
  Collaboration}(2011)}]{Wobble2011arXiv1111.0121F}
{Finnegan}, G., \& {for the VERITAS Collaboration}. 2011, ArXiv e-prints.
\newblock \doarXiv{1111.0121}

\bibitem[{{Gilmore} {et~al.}(2012){Gilmore}, {Somerville}, {Primack}, \&
  {Dom{\'{\i}}nguez}}]{Gilmore}
{Gilmore}, R.~C., {Somerville}, R.~S., {Primack}, J.~R., \& {Dom{\'{\i}}nguez},
  A. 2012, Mon. Not. R. Astron Soc, 422, 3189,
  \dodoi{10.1111/j.1365-2966.2012.20841.x}

\bibitem[{{Gould} \& {Rephaeli}(1978)}]{1978ApJ...225..318G}
{Gould}, R., \& {Rephaeli}, Y. 1978, Astrophys. J., 225, 318,
  \dodoi{10.1086/156493}

\bibitem[{{Gould} \& {Schr{\'e}der}(1966)}]{1966PhRvL..16..252G}
{Gould}, R., \& {Schr{\'e}der}, G. 1966, Phys. Rev. Lett., 16, 252,
  \dodoi{10.1103/PhysRevLett.16.252}

\bibitem[{{H.E.S.S.~Collaboration} {et~al.}(2010){H.E.S.S.~Collaboration},
  {Abramowski}, {Acero}, {Aharonian}, {Akhperjanian}, \&
  {Anton}}]{2010A&A...520A..83H}
{H.E.S.S.~Collaboration}, {Abramowski}, A., {Acero}, F., {et~al.} 2010, \aap,
  520, A83, \dodoi{10.1051/0004-6361/201014484}

\bibitem[{{Kachelrie{\ss}} {et~al.}(2012){Kachelrie{\ss}}, {Ostapchenko}, \&
  {Tom{\`a}s}}]{Elmag2012CoPhC.183.1036K}
{Kachelrie{\ss}}, M., {Ostapchenko}, S., \& {Tom{\`a}s}, R. 2012, Computer
  Physics Communications, 183, 1036, \dodoi{10.1016/j.cpc.2011.12.025}

\bibitem[{{Kotelnikov} {et~al.}(2015){Kotelnikov}, {Rubtsov}, \&
  {Troitsky}}]{2015MNRAS.450L..44K}
{Kotelnikov}, E., {Rubtsov}, G., \& {Troitsky}, S. 2015, \mnras, 450, L44,
  \dodoi{10.1093/mnrasl/slv044}

\bibitem[{{Meyer} {et~al.}(2016){Meyer}, {Conrad}, \&
  {Dickinson}}]{Meyer2016ApJ...827..147M}
{Meyer}, M., {Conrad}, J., \& {Dickinson}, H. 2016, Astrophys. J., 827, 147,
  \dodoi{10.3847/0004-637X/827/2/147}

\bibitem[{{Neronov} \& {Semikoz}(2009)}]{Neronov2009PhRvD..80l3012N}
{Neronov}, A., \& {Semikoz}, D.~V. 2009, \prd, 80, 123012,
  \dodoi{10.1103/PhysRevD.80.123012}

\bibitem[{{Pshirkov} {et~al.}(2015){Pshirkov}, {Tinyakov}, \&
  {Urban}}]{2015MNRAS.452.2851P}
{Pshirkov}, M.~S., {Tinyakov}, P.~G., \& {Urban}, F.~R. 2015, Mon. Not. R.
  Astron Soc, 452, 2851, \dodoi{10.1093/mnras/stv1273}

\bibitem[{{Takahashi} {et~al.}(2011){Takahashi}, {Inoue}, {Ichiki}, \&
  {Nakamura}}]{2011MNRAS.410.2741T}
{Takahashi}, K., {Inoue}, S., {Ichiki}, K., \& {Nakamura}, T. 2011, \mnras,
  410, 2741, \dodoi{10.1111/j.1365-2966.2010.17639.x}

\bibitem[{{Vincent} \&
  {H.E.S.S.~Collaboration}(2004)}]{HESS2004sf2a.conf..407V}
{Vincent}, P., \& {H.E.S.S.~Collaboration}. 2004, in SF2A-2004: Semaine de
  l'Astrophysique Francaise, ed. F.~{Combes}, D.~{Barret}, T.~{Contini},
  F.~{Meynadier}, \& L.~{Pagani}, 407

\bibitem[{{Wakely} \& {Horan}(2008)}]{2008ICRC....3.1341W}
{Wakely}, S.~P., \& {Horan}, D. 2008, International Cosmic Ray Conference, 3,
  1341

\end{thebibliography}

%\begin{acknowledgements}
%A.~D. Supanitsky and A. Rovero are members of the {\it Ca\-rre\-ra del Investigador Cient\'{\i}fico} of CONICET, Argentina.
%\end{acknowledgements}

\end{document}